\documentclass[aps,jcp,amsmath,amssymb,reprint]{revtex4-2}

\usepackage{chemformula} % Formula subscripts using \ch{}
\usepackage[T1]{fontenc} % Use modern font encodings
\usepackage{multirow}
\usepackage{dcolumn}
\usepackage{tabularx}
\usepackage{float}
\usepackage{xr}
\usepackage{hyperref}
\usepackage[utf8]{inputenc}
\usepackage{mciteplus}
\usepackage{subfig}
\usepackage[utf8]{inputenc}
\setcitestyle{super}
\usepackage[cdot,mediumqspace,amssymb]{SIunits}
\usepackage{amsmath}

\usepackage{textcomp} % to get a tilde
\newcommand{\comment}[1]{}

\begin{document}
	\preprint{AIP/123-QED}
	
	\title{Detection of infrared light through stimulated four-wave mixing process}
	
	\author{Wei-Hang Zhang}
	\affiliation{CAS Key Laboratory of Quantum Information, University of Science and Technology of China, Hefei 230026, China}
	\affiliation{CAS Center for Excellence in Quantum Information and Quantum Physics, University of Science and Technology of China, Hefei 230026, China}
		
	\author{Jing-Yuan Peng}
	\affiliation{CAS Key Laboratory of Quantum Information, University of Science and Technology of China, Hefei 230026, China}
	\affiliation{CAS Center for Excellence in Quantum Information and Quantum Physics, University of Science and Technology of China, Hefei 230026, China}
	
	\author{En-Ze Li}
	\affiliation{CAS Key Laboratory of Quantum Information, University of Science and Technology of China, Hefei 230026, China}
	\affiliation{CAS Center for Excellence in Quantum Information and Quantum Physics, University of Science and Technology of China, Hefei 230026, China}
	
	\author{Ying-Hao Ye}
	\affiliation{CAS Key Laboratory of Quantum Information, University of Science and Technology of China, Hefei 230026, China}
	\affiliation{CAS Center for Excellence in Quantum Information and Quantum Physics, University of Science and Technology of China, Hefei 230026, China}
	
	\author{Lei Zeng}
	\affiliation{CAS Key Laboratory of Quantum Information, University of Science and Technology of China, Hefei 230026, China}
	\affiliation{CAS Center for Excellence in Quantum Information and Quantum Physics, University of Science and Technology of China, Hefei 230026, China}
	
	\author{Dong-Sheng Ding}
	\email{dds@ustc.edu.cn}
	\affiliation{CAS Key Laboratory of Quantum Information, University of Science and Technology of China, Hefei 230026, China}
	\affiliation{CAS Center for Excellence in Quantum Information and Quantum Physics, University of Science and Technology of China, Hefei 230026, China}
	
	\author{Bao-Sen Shi}%
	\email{drshi@ustc.edu.cn}
	\affiliation{CAS Key Laboratory of Quantum Information, University of Science and Technology of China, Hefei 230026, China}
	\affiliation{CAS Center for Excellence in Quantum Information and Quantum Physics, University of Science and Technology of China, Hefei 230026, China}

\begin{abstract}

Infrared optical measurement has a wide range of applications in industry and science, but infrared light detectors suffer from high costs and inferior performance than visible light detectors. Four-wave mixing (FWM) process allows detection in the infrared range by detecting correlated visible light. We experimentally investigate the stimulated FWM process in a hot $^{85}$Rb atomic vapor cell, in which a weak infrared signal laser at $1530~$nm induces the FWM process and is amplified and converted into a strong FWM light at $780~$nm, the latter can be detected more easily. We find the optimized single- and two-photon detunings by studying the dependence of the frequency of input laser on the generated FWM light. What's more, the power gain increases rapidly as the signal intensity decreases, which is consistent with our theoretical analysis. As a result, the power gain can reach up to 500 at a signal laser power of $0.1~\mu$W and the number of detected photons increased by a factor of 250. Finally, we experimentally prove that our amplification process can work in a broad band in the frequency domain by exploring the response rate of our stimulated FWM process.

\end{abstract}
\maketitle


\section{Introduction}

Infrared optical measurement is an important technique for applications in scientific research and industrial production. However, infrared light detectors in infrared optical devices have higher dark count rates, lower specific detectivities, and higher cost than the detector in the visible range, and often need to be cooled \cite{b1}. For this dilemma, researchers correlate infrared light with visible light through nonlinear optical phenomena, and measure infrared light by detecting visible light with visible light detectors \cite{a1,a2}. Four-wave mixing (FWM) is a common method for generating high-power visible light \cite{a40} and converting infrared light to visible light \cite{a29,a30,a31}, the latter is investigated in our work.

Four-wave mixing process is one of the prominent nonlinear optical phenomena \cite{a3} and originates from the third-order optical nonlinear effects in the interaction of optical fields with a nonlinear medium. In atomic system, the FWM process has been explored extensively due to its wide applications such as observing optical parametric amplification in the process of degenerate stimulated four-photon interaction \cite{a4} and optical precursors phenomena \cite{a32}; generating  non-classical correlated photon pairs \cite{a5,a6,a33,a36} via the spontaneous FWM process for use in long-distance quantum communication \cite{a7} and optical quantum information processing \cite{a9}; observing squeezed light \cite{a10} and strong relative intensity squeezing \cite{a11}; exhibiting localized entanglement of twin images \cite{a12}; and converting frequency of the optical field \cite{a13,a14}. The FWM process has been realized in atomic system with various energy-level configurations such as double-ladder type \cite{a15,a16}, double-lambda type \cite{a17,a18,a34}, diamond type \cite{a19,a20} and inverted-Y type \cite{a21}. In this paper, we focus on a four-level system with a diamond-type configuration.

The stimulated FWM process with a four-level atomic system is an extensively researched topic in a double-ladder or diamond configuration. This phenomenon requires the interaction of three optical fields with the atoms in which two strong fields excite the atoms to a higher excited state (such as 5D or 4D state in rubidium atomic system) and a weak seed (driving) field induces the atoms to generate a strong FWM field. This process is also referred to as seeded FWM in some works \cite{a22}. To date, the stimulated FWM process in a ladder-type atomic system has been studied to investigate the effects of atomic coherence on the FWM spectrum \cite{a23} or the relationship between three-photon electromagnetically induced absorption and FWM \cite{a24}. The biphoton spectral wave form of photon pairs emitted from a cascade-type atomic ensemble also can be measured by using the stimulated FWM process \cite{a25}. Recently, the amplified spontaneous emission induced self-stimulated FWM process and the stimulated FWM process have been exploited in the diamond-type atomic system \cite{a22}, and the stimulated nondegenerate FWM phenomenon without amplified spontaneous emission has been realized in cesium atoms with a power of the generated FWM light up to $1.2~$mW \cite{a26}. The stimulated FWM process has also been investigated in rubidium atoms via Rydberg states \cite{a27}. To put it briefly, the stimulated FWM is generally utilized to explore precision spectroscopic measurements, atomic coherence, and efficient generation of FWM light.

\begin{figure*}[htp!]
	\begin{center}
		\includegraphics[width=17.5cm]{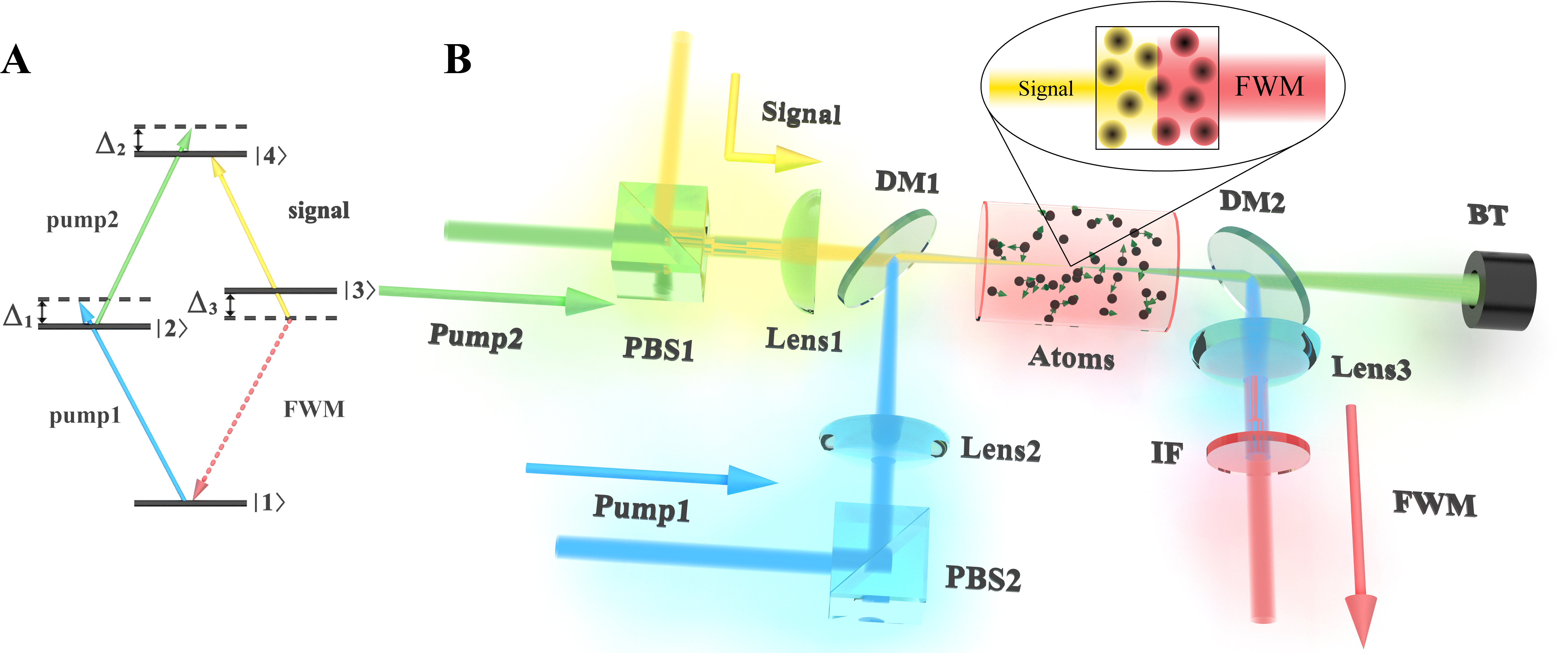}% This is a *.eps file
	\end{center}
	\caption{(color online) (A) Energy-level diagram of diamond configuration. (B) Schematic diagram of the experimental setup. PBS1, PBS2: polarizing beam splitter; DM: dichroic mirror; Lens: planoconvex lens (the focal length of Lens1, Lens2 and Lens3 are 150~mm, 300~mm and 150~mm respctively); IF: Interference filter; BT: beam traps.}\label{fig:1}
\end{figure*}

\begin{figure*}[htp!]
	\begin{center}
		\includegraphics[width=17cm]{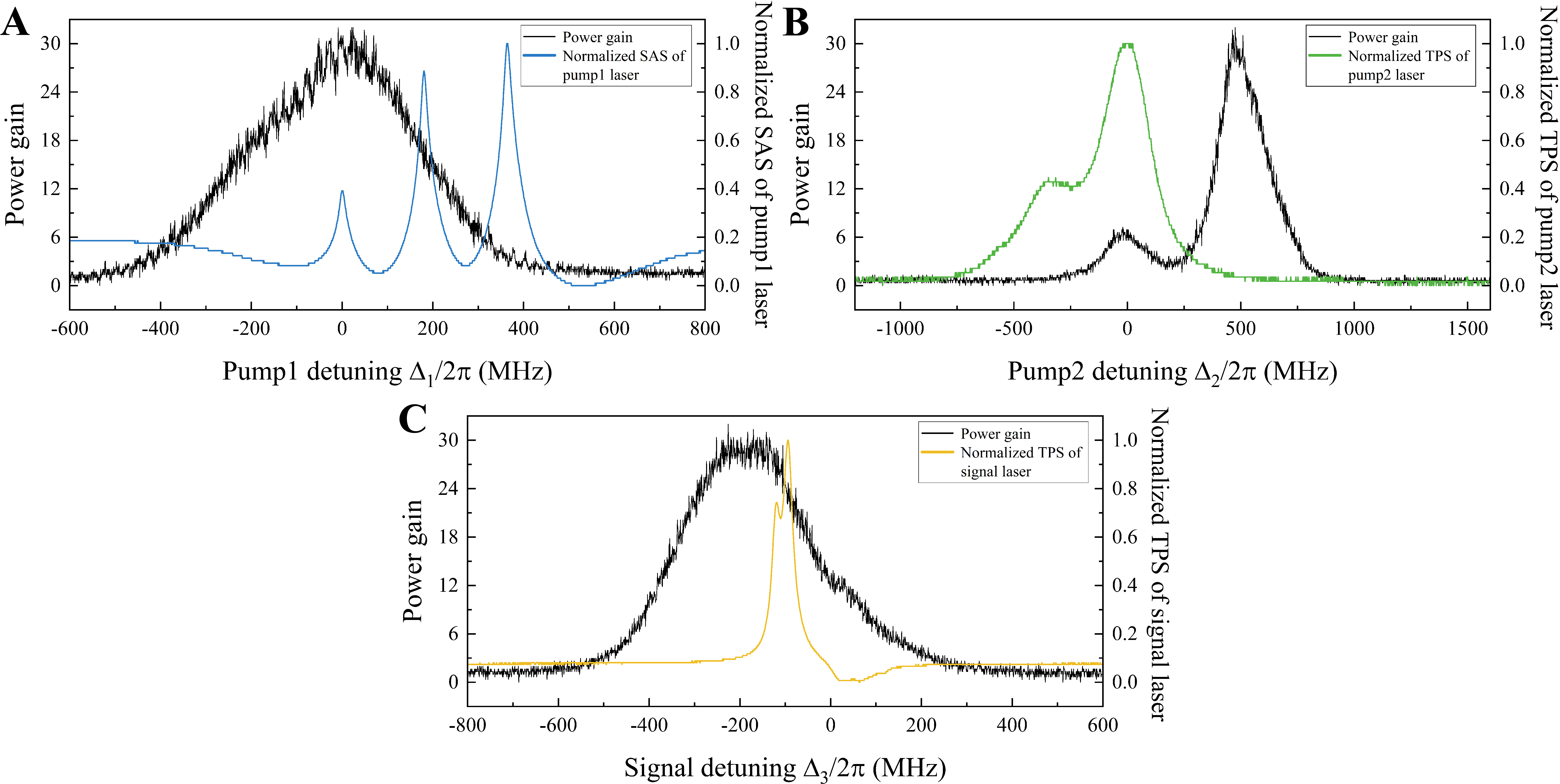}
	\end{center}
	\caption{(color online) (A) Power gain (black curve) and normalized SAS of the pumpl laser (blue curve) against the detuning of the pump1 laser. (B) Power gain (black curve) and normalized SAS of the pump2 laser (green curve) against the detuning of the pump2 laser. (C) Power gain (black curve) and normalized TPS of the signal laser (yellow curve) against the detuning of the signal laser.}\label{fig:2}
\end{figure*}

In this paper, we report on an experimental exploration of the stimulated FWM process in a diamond-type $^{85}$Rb atomic ensemble. Realizing the amplified detection of the infrared laser by converting the weak infrared signal light into the strong visible FWM light. In this work, the atoms are excited and populated in the upper state by inputting two strong pump lasers and we make an infrared beam as a seed light to induce atoms to efficiently generate the FWM light. Specifically, we find that we can achieve the purpose of amplifying and detecting the infrared signal by detecting the strong visible FWM signal which is induced by this infrared signal. In addition, in order to investigate the response properties of the amplifier, we compare the waveforms between the input signal and the output signal for various rising edge cases.

\section{Experimental setup}

We experimentally realized the amplified detection of infrared signal through a stimulated FWM process in hot $^{85}$Rb atomic vapor with a four-levels diamond configuration as shown in Figure \ref{fig:1}A. It consists of one ground state $|1\rangle$ ($5S_{1/2}(F=2)$), one excited state $|4\rangle$ ($4D_{3/2}(F^{\prime\prime}=3)$), and two intermediate states $|2\rangle$ and $|3\rangle$ ($5P_{1/2}(F^{\prime}=2)$ and $5P_{3/2}(F^{\prime}=3)$). The pump1 ($795~$nm) and pump2 ($1475~$nm) lasers excite the atoms from $|1\rangle$ to $|2\rangle$ and $|2\rangle$ to $|4\rangle$ respectively, then the atoms decay into the ground state $|1\rangle$ through the intermediate state $|3\rangle$. Since the difference between the spontaneous decay rates ($\Gamma_{43}$) from $|4\rangle$ to $|3\rangle$ and the spontaneous decay rates ($\Gamma_{31}$) from $|3\rangle$ to $|1\rangle$ satisfies the relation $\Gamma_{43}=0.1\Gamma_{31}$, the populations of $|4\rangle$ and $|3\rangle$ are not significantly inversed and the FWM process cannot occur obviously in the absence of the signal laser \cite{a22}. So that we input the signal ($1530~$nm) laser which interacts with the $|3\rangle-|4\rangle$ transition to induce the FWM, and the FWM ($780~$nm) light can be generated strongly in the transition from $|3\rangle$ to $|1\rangle$ under the phase-matching condition of wave-vector conservation and energy conservation. Through this stimulated FWM process, a weak signal laser can be converted into a strong FWM light to achieve the amplification of the infrared signal laser. The detuning frequencies of the pump1, pump2, and signal lasers correspond to $\Delta_{1}$, $\Delta_{2}$ and $\Delta_{3}$ respectively.

The experimental setup is schematically depicted in Figure \ref{fig:1}B, where pump 1, pump2 and signal lasers are continuous waves (CW) and overlap in a 5-cm-long $^{85}$Rb cell that is heated to 140 $^{\circ}$C. The horizontally polarized pump2 beam and vertically polarized signal beam are focused into the center of the cell with the same diameter of $126~\mu$m. The vertically polarized pump1 beam is superimposed with the pump2 beam and the signal beam through a dichromatic mirror (DM1), and has a diameter of $141~\mu$m in the focal spot. The strong pump2 laser and signal laser are filtered out by the DM2, and the pump1 laser is blocked by a subsequent interference filter (IF). A photomultiplier tube (PMT) with an adjustable attenuation collects the generated FWM light in the direction which is colinear with the pump1 laser in our system. 

\begin{figure}[h!]
	\begin{center}
		\includegraphics[width=8cm]{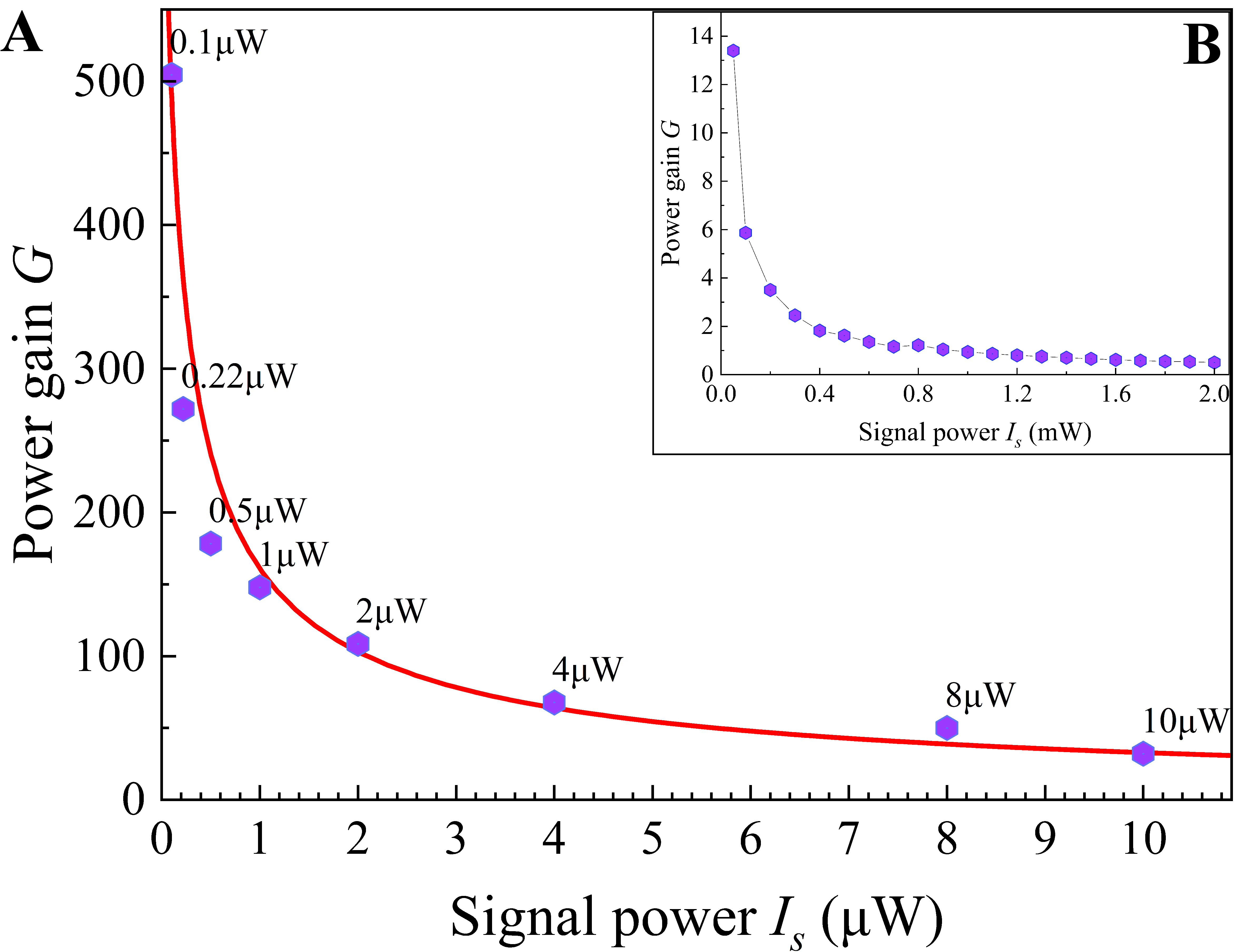}
	\end{center}
	\caption{(color online) Power gain $G$ between FWM light and signal laser against the power of the signal laser. (A) Experimental datas (purple dots) and theoretical fitting (red curve) of signal power $I_{s}$ from $0.1~\mu$W to $10~\mu$W. (B) Experimental datas (purple dots) of signal power $I_{s}$ from $0.05~$mW to $2~$mW.}\label{fig:3}
\end{figure}

\section{Experimental results and theoretical analysis}

A critical metric to characterize the performance of our system is the power gain ($G=I_{F}/I_{s}$) defined as the ratio of the FWM light power ($I_{F}$) and the signal laser power ($I_{s}$). Figure \ref{fig:2} shows the variation of the power gain (three black curves) with the detuning frequencies of pump1, pump2, and signal lasers ($\Delta_{1}$, $\Delta_{2}$ and $\Delta_{3}$), where the powers of the signal, pump1 and pump2 lasers are $10~\mu$W, $110~$mW and $110~$mW respectively. In Figure \ref{fig:2}A, the power gain are measured when we scan the detuning frequency of pump1 $\Delta_{1}$ laser and the detuning frequency of pump2 $\Delta_{2}$ (signal $\Delta_{3}$) laser is fixed at $475\times2\pi~$MHz ($-190\times2\pi~$MHz). The maximum power gain approaches 30 when the $\Delta_{1}$ is equal to $0\times2\pi~$MHz, and the blue curve is the saturated absorption spectrum (SAS) of the pump1 field. We can find that the power gain has two peaks when the pump1 laser is resonant with the $|1\rangle-|2\rangle$ transition and the frequency of the signal laser is satisfied with $\Delta_{3}=-190\times2\pi~$MHz, as shown in Figure \ref{fig:2}B. It's due to another two peaks which are corresponded to the transitions of $5S_{1/2}(F=2)-5P_{1/2}(F^{\prime}=3)-4D_{3/2}(F^{\prime\prime}=3)$ and $5S_{1/2}(F=2)-5P_{1/2}(F^{\prime}=2)-4D_{3/2}(F^{\prime\prime}=3)$ in the normalized two photons spectrum (TPS) of pump2 laser (green curve) and the maximum power gain occurs at the condition of $\Delta_{2}=475\times2\pi~$MHz. Figure \ref{fig:2}C displays the power gain and the normalized TPS of the signal laser (yellow curve) when we scan the $\Delta_{3}$ of the signal laser under the conditions of $\Delta_{1}=0\times2\pi~$MHz and $\Delta_{2}=475\times2\pi~$MHz. It is obvious from Figure \ref{fig:2} that the single-photon detunings of pump and signal fields significantly affect the value of the power gain. The maximum point of the power gain reaches 30 when the $\Delta_{1}$, $\Delta_{2}$, and $\Delta_{3}$ are fixed at $0\times2\pi~$MHz, $475\times2\pi~$MHz, and $-190\times2\pi~$MHz respectively.

\begin{figure*}[htp!]
	\begin{center}
		\includegraphics[width=17.5cm]{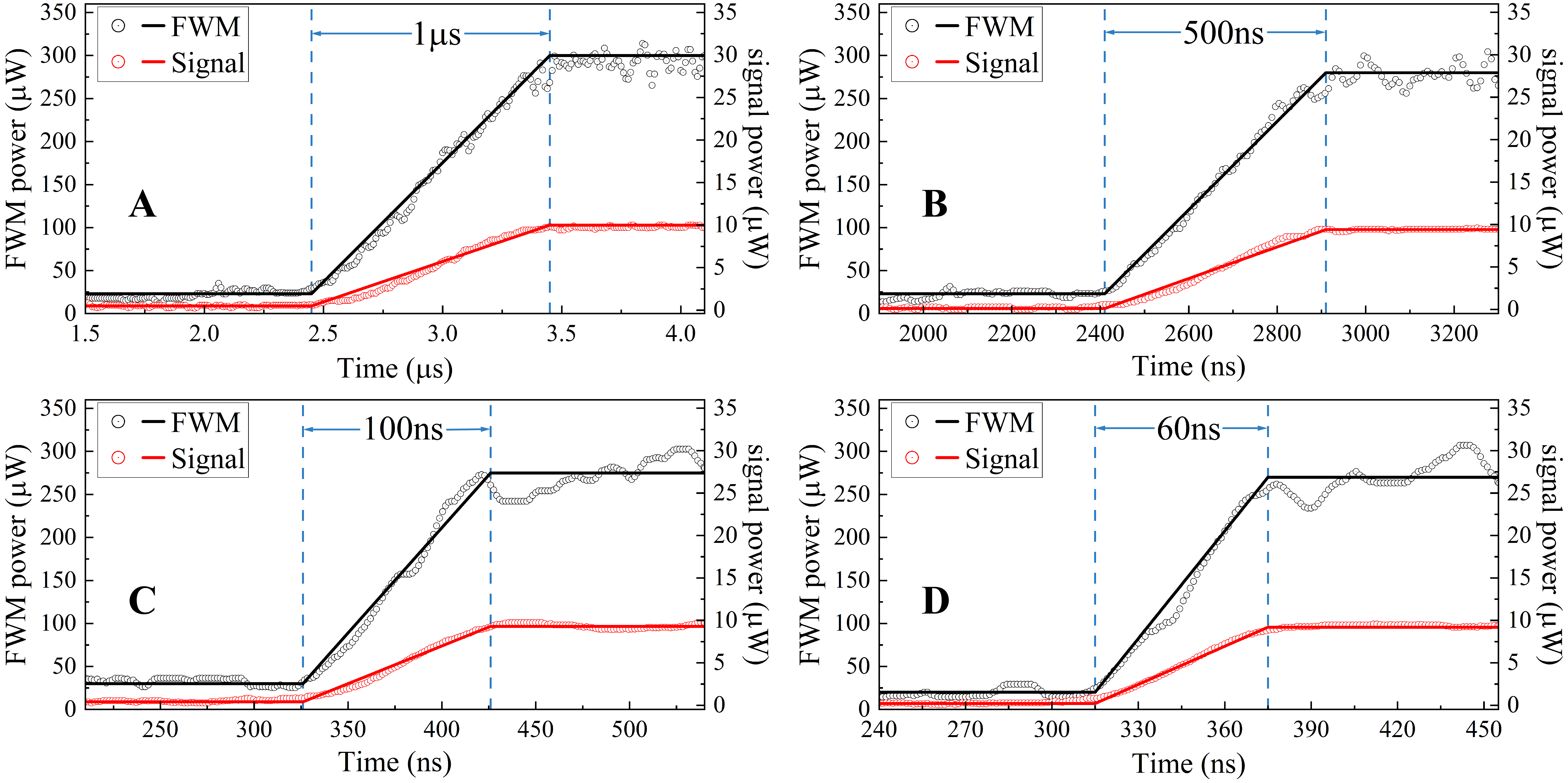}
	\end{center}
	\caption{(color online) The measured wavefroms of the input pulsed infrared signal laser (red curve) and the output pulsed FWM laser (black curve) with a rising time of (A) $1~\mu$s, (B) $500~$ns, (C) $100~$ns, (D) $60~$ns.}\label{fig:4}
\end{figure*}

In Figure \ref{fig:3}, we plot the power gain as a function of the signal power $I_{s}$ under the optimal experimental conditions we mentioned above, and the power is measured with a power meter instead of the PMT. Due to the strong pump1 and pump2 lasers, the atoms are mostly populated at the state $|4\rangle$ via two-photon excitation. Meanwhile, the weak signal laser is converted and amplified into a strong FWM light. We can fetch the information from Figure \ref{fig:3} that the power gain increases rapidly with the decrease of the signal power. Figure \ref{fig:3}(a) indicates that the power gain exceeds 100 when the power of the signal laser is less than $2~\mu$W, and the power gain can be reached up to 500 in the case of $I_{s}=0.1~\mu$W. The amplification effect exhibits a tendency to saturation when the signal power intensity is beyond $1~$mW, which is illustrated in Figure \ref{fig:3}B. We also can defined the photon-number gain (or detection gain) $G_{photon}$ in this amplification process as the ratio of the number of output visible photons to the number of input infrared photons. Considering that the energy of a photon is inversely proportional to its wavelength, we can obtain the relationship $G_{photon}=0.5G$ between photon-number gain and power gain. Similarly, the photon-number gain can reach 250 at the case of $I_{s}=0.1~\mu$W according to Figure \ref{fig:3}B, which means that the number of photons detected by a detector is increased from 1 to 250.

To explain the amplification of signal laser in the stimulated FWM process, we theoretically calculate the relation between power gain and signal power in an atomic diamond-type energy-level system. The equation of motion can be described as \cite{b3}
\begin{equation}
\label{eq:01}
\frac{\partial \rho_{ij}}{\partial t}=-\frac{i}{\hbar}\sum_{k}{\left[H_{i k}\rho_{kj}-\rho_{i k}H_{kj}\right]}-\Gamma_{ij}\rho_{ij}-\gamma_{ij}\rho_{ij}
\end{equation} 
where $\rho_{ij}$ is the density-matrix element and $H_{ij}$ is the effective interaction Hamiltonian. The subscript indices $i$ and $j$ indicate the $|i\rangle$ and $|j\rangle$ states respectively. $\Gamma_{ij}$ denotes the population spontaneous relaxation rate from state  $|i\rangle$ to state $|j\rangle$ while $\gamma_{ij}=(\Gamma_{ii}+\Gamma_{jj})/2$ denotes the relaxation rates of the coherences.  For $^{\mathrm{85}}\mathrm{Rb}$, the spontaneous decay rates are $\Gamma_{33}=6\times2\pi~$MHz, $\Gamma_{22}=\Gamma_{33}$, $\Gamma_{44}=0.35\Gamma_{33}$, $\Gamma_{43}=0.1\Gamma_{33}$, and $\Gamma_{42}=0.25\Gamma_{33}$. The matrix of the  effective interaction Hamiltonian  $\hat{H}$ can be represented as 
\begin{equation}
\label{eq:02}
\hat{H}=\frac{\hbar}{2}\left(\begin{matrix}
0& \Omega^{*}_{21} & 0 & 0 \\ 
\Omega_{21}& -2\Delta_{1} & 0 & \Omega^{*}_{42} \\ 
0& 0 & -2\Delta_{4} & \Omega^{*}_{43} \\ 
0& \Omega_{42} & \Omega_{43} & -2(\Delta_{1}+\Delta_{2})
\end{matrix}\right)
\end{equation} 
with $\Delta_{4}=\Delta_{1}+\Delta_{2}-\Delta_{3}$, where $\Omega_{21}$, $\Omega_{42}$ and $\Omega_{43}$ are the Rabi frequency of pump1, pump2, and signal laser respectively. We can then deduce the steady-state optical coherence $\rho_{13}$ from equation (\ref{eq:01}) and equation (\ref{eq:02}) as \cite{a37}:
\begin{equation}
\label{eq:03}
\rho_{31}=\frac{-i\Omega_{21}\Omega_{42}\Omega^{*}_{43}}{[2i\Delta_{1}-\gamma_{21}][2i(\Delta_{1}+\Delta_{2})-\gamma_{41}][2i\Delta_{4}-\gamma_{31}]}.
\end{equation}
The amplitude of the FWM light wave $E_{FWM}$ is proportional to the third-order optical polarization $P^{(3)}_{FWM}$ \cite{a28}:
\begin{equation}
\label{eq:04}
E_{FWM}\propto P^{(3)}_{FWM}=N\mu^{*}_{31}\rho_{31}+\mathrm{H.c}.
\end{equation}
where $N$ is the number density of atoms and $\mu^{*}_{31}$ is the transition dipole moment between states $|1\rangle$ and $|3\rangle$. In our system, the third-order optical polarization can be represented as $E_{FWM}\propto P^{(3)}_{FWM}=\epsilon_{0}\chi^{(3)}E_{p1}E_{p2}E^{*}_{s}$, where $\chi^{(3)}$ denotes the third-order susceptibility; $\epsilon_{0}$ is the vacuum permittivity; $E_{p1}$, $E_{p2}$ and $E_{s}$ are amplitudes of the pump1, pump2 and the signal electric fields, respectively. We can obtain the power conversion efficiency $\eta\propto E^{2}_{FWM}/E^{2}_{s}$ between signal light and FWM light by calculating $\chi^{(3)}$. 

Considering the stimulated amplified process of signal light, we treat the state $|3\rangle$ and the state $|4\rangle$ equivalents as a simple two-level atomic system for the signal field. The stimulated emission from the interaction between this two-level atomic system and signal field can enhance the input signal light with a gain factor of $G_{factor}=I_{sout}/I_{s}$ ($I_{sout}$ is the power of the amplified signal light), and $G_{factor}$ is satisfied by the following equation which is described in references \cite{a38,b2}: 
\begin{equation}
\label{eq:05}
I_{s}(G_{factor}-1)+2A\ln G_{factor}=\frac{1}{2}B.
\end{equation}
with $A=\pi\omega^{2}_{s}I_{sat}\frac{\Delta^{2}_{3}+\Gamma^{2}_{43}}{\Gamma^{2}_{43}}$ and $B=\alpha N\pi\omega^{2}_{s}h\nu L$, where $\omega_{s}$ is the waist of signal light, $\nu$ is the frequency of the signal light at 1530nm, $L$ is the length of the $^{85}$Rb vapor cell, $I_{sat}=\pi hc\Gamma_{43}/(3\lambda^{3})$ is the saturation intensity, and $\alpha=1/\tau_{cyc}$ represents the pumping rate with $\tau_{cyc}=\frac{1}{\Omega_{43}}+\frac{1}{\Gamma_{43}}+\frac{1}{\Gamma_{31}}$ being the atomic cycling time \cite{a39}. We obtain the power gain 
\begin{equation}
\label{eq:06}
G=\eta G_{factor}
\end{equation}
by considering that the stimulated FWM process consists of an amplification process and a frequency conversion process, where $\eta$ is a constant and $G_{factor}$ is a function of $I_{s}$. We then fit our experimental data with the deduced function and the result is shown in Figure \ref{fig:3}A. The theoretical analysis agrees with our experimental results.

Finally, for amplified detection of an infrared signal, we should also pay attention to the response rate in this conversion process. As illustrated in Figure \ref{fig:4}, we replace the CW signal laser with a pulsed signal laser. Figure \ref{fig:4}A-D compare the signal pulse with the generated FWM pulse at a rising time of $1~\mu$s, $500~$ns, $100~$ns and $60~$ns, respectively. It can be concluded that the rising time of the input signal light and output FWM light are almost identical and the response time of this conversion process is faster than $60~$ns. The study of higher response rate is mainly restricted by the shortest achievable rising time of the acoustic optical modulator(AOM) in our setup, which is $55~$ns.

\section{Conclusion}

In conclusion, we have demonstrated the amplified detection of an infrared signal via a stimulated FWM process based on an atomic diamond-type configuration. What's more, the generated FWM light is in the visible range, which can be detected more efficiently. We find that the power gain increases sharply as the power of the signal laser decreases, and the experimental data is in good agreement with our theoretical fitting. We also investigate the dependence of power gain on the detunings of the three input lasers. Simultaneously, we experimentally confirm that the response rate of the FWM process is faster than $60~$ns, therefore it can be applied to the detection of the weak infrared signals in specific situations.

\section*{Data availability statement}
The raw data supporting the conclusion of this article will be made available by the authors, without undue reservation.

\section*{Author contributions}
D-SD and B-SS coordianted the research project. W-HZ peformed the experimental fabrication, measurments and analyzed the data. All authors discussed the manuscript.

\section*{Conflict of interest}
The authors declare that the research was conducted in the absence of any commercial or financial relationships that could be construed as a potential conflict of interest.

\section*{Funding}
We acknowledge funding from National Natural Science Foundation of China (Grant Nos. U20A20218, 61525504, and 11934013), Youth Innovation Promotion Association of the Chinese Academy of Sciences (Grant No. 2018490) and the Major Science and Technology Projects in Anhui Province (Grant No. 202203a13010001).

\bibliographystyle{jcp}
\bibliography{Hotatoms}

\end{document}